\documentclass[pmlr]{jmlr}

\newcommand{\so}[1]{{\color{green} {\textbf{sageev: #1}}}}
\newcommand{\socheck}[1]{{\color{blue} {\textbf{sageev (note to self to check this):} {\textit{#1}}}}}

\newcommand{\hide}[1]{}

\jmlrworkshop{NeurIPS 2020 Competition and Demonstration Track}

\title[Musical Speech]{Musical Speech: A Transformer-based Composition Tool}

  \author{\Name{Jason d'Eon}\thanks{Equal Contribution} \Email{jndeon@dal.ca}\\
  \Name{Sri Harsha Dumpala}\textsuperscript{*} \Email{sriharsha.d@dal.ca}\\
  \Name{Chandramouli Shama Sastry}\textsuperscript{*} \Email{cssastry@dal.ca}\\
  \addr Dalhousie University, Vector Institute
  \AND
  \Name{Dani Oore} \Email{doore@mun.ca}\\
  \addr IICSI, Memorial University of Newfoundland
  \AND
  \Name{Sageev Oore} \Email{sageev@vectorinstitute.ai}\\
  \addr Dalhousie University, Vector Institute
 }

\editors{Hugo Jair Escalante and Katja Hofmann}

\begin{document}

\maketitle

\begin{abstract}
In this paper, we propose a new compositional tool that will generate a musical outline of speech recorded/provided by the user for use as a musical building block in their compositions. The tool allows any user to use their own speech to generate musical material, while still being able to hear the direct connection between their recorded speech and the resulting music. The tool is built on our proposed pipeline. This pipeline begins with speech-based signal processing, after which some simple musical heuristics are applied, and finally these pre-processed signals are passed through Transformer models trained on new musical tasks.
We illustrate the effectiveness of our pipeline -- which does not require a paired dataset for training -- through examples of music created by musicians making use of our tool.
\end{abstract}
\begin{keywords}
Speech processing, musical notes, transformer networks, denoising autoencoder.
\end{keywords}


\section{Introduction}

Among the extensive recent works applying machine learning to the generation of music and sound \citep{dhariwal2020jukebox, dieleman2018challenge, aiva, payne2019musenet,engel2020ddsp,oore2018time, vasquez2019melnet, herremans2017taxonomy}, an interesting and important direction are those works that aim to provide musicians with compositional tools. One approach in this domain is tools that ``generate material'', perhaps in some way based on initial musical ideas or directions set by the user, e.g. \cite{huang2017counterpoint, huang2018music, donahue2019,roberts2018learning, meade2019exploring, sturm2019machine}. Of particular interest to us are vocal-based tools that generate musical ideas conditioned on a recording of the user's voice. The concept of converting speech into music has a long musical history (see Section~\ref{s:background-s2m} for a detailed discussion), but no techniques to automate parts of the process have been described in the literature as far as we know. It is an important compositional technique that is challenging to do even for highly skilled musicians, yet the results have often been very effective. While we are not aware of studies that explore why, in our experience, musicians who have a skilled enough ear (and/or perfect pitch) that they are able to accurately transcribe musical excerpts, still find it challenging and time-consuming to ``transcribe'' speech to music.



There is another, quite different reason we are interested in converting speech to music: separating prosody and timbre from content provides an informative sonification of paralinguistic characteristics of speech (e.g. \cite{harsha2020}), which are informative distinctly from the semantic content. For example, recent work uses paralinguistic characteristics of speech to predict information about the mental health of the speaker~\citep{harsha2021, cheng2020advances}. In another example, musically-driven analyses of rhythmic aspects of speech (obtained by meticulous manual extraction of musical characteristics from political speeches) have been used to discover rhythmic motifs and examine their role in socio-political contexts~\citep{oore2018trump}.

We thus propose developing machine learning tools that allow a user to generate musical source material by translating their speech into melodic fragments. To achieve this we combine speech-based signal processing, musical heuristics, and a Transformer model. We needed to define new musical tasks in order to train this Transformer model, that would represent the requirements of our context. Specifically, the raw outputs of the speech processing (e.g. extraction of formant parameters, as described below) result in a barrage of quickly changing musical pitches, and we needed a system that could find a medium between this raw output, and pure generative model outputs.

We will outline the approach in Section~\ref{s:contributions}, provide details in the next sections, and in Section~\ref{s:evaluation} present some analysis of the results and, crucially, examples of compositional excerpts created from short audio recordings of speech using our system. First we give background information on the three main elements of this system: speech processing, conditional music generation, and the historical conversion of speech into music.

\section{Background and Related Work}
\label{s:background}

\subsection{Speech Processing}
\label{s:background-speech}


Speech and music are two distinct aural phenomena perceived by the same human auditory mechanism, and often treated separately (e.g. whether for analysis or generation). A variety of works have explored some of the connections between these phenomena \citep{hausen2013music, ding2017temporal}. \cite{ding2017temporal} show that the rhythmic structure is a fundamental feature of both speech and music. Both domains involve sequences of events (such as syllables or notes) which have systematic patterns of timing, accent, and grouping \citep{patel2010music}, but the rhythmic pattern of speech is distinct from that of music \citep{ding2017temporal}. In this paper, we slightly adjust rhythmic patterns of speech to map them to rhythmic patterns of music.
Further, it was observed by linguists that
human speech has a temporal rhythm that can be characterized by placing a perceptual “beat” around successive syllables \citep{port2003meter}.
In this work, we exploit this observation to obtain an intermediate representation of the speech signal by considering the acoustic features around regions with high intensity or around syllable nuclei positions in speech.


\subsection{Music Generation}
\label{s:background-music-generation}


In recent years, considerable work has been done in the application of machine learning in musical contexts. Such musical generation and processing is generally approached at one of two levels: (1) at the raw audio level, where a waveform is represented at a sample rate on the order of 16 KHz, and (2) at the note-level, where the gaps between notes of a musical performance can be represented with a sample rate of roughly 50Hz (that is, there are usually no more than 16 notes per second, usually much fewer, but for reasonable fidelity, the spaces between them need to be represented with a resolution of at least 20ms or better). The note-level approach---and the one that we take here---is considered symbolic (i.e. each ``note'' is one element in a finite set of notes, rather than a continuous waveform), and the representation is usually based on the MIDI-format. For purposes of this paper, this simply means each note is a symbol, and each durational event (e.g. length of a note, distance between onsets of two notes) has been discretized and can thus be tokenized as well.

In general, generating symbolic music \citep{ huang2018music,payne2019musenet,oore2018time} in the form of MIDI is relatively easier than directly generating raw audio of music \citep{dhariwal2020jukebox,dieleman2018challenge, vasquez2019melnet} as generating raw audio involves maintaining coherence over several thousand samples even for a clip of few milliseconds. We use speech processing techniques to move from the waveform domain of raw speech to the symbolic domain of note-based representation, and then build upon previous works~\cite{huang2018music,oore2018time} to transform one (speech-derived) note sequence into another (musical) one. 


\hide{In addition to unguided music generation, much work is being done towards creating tools which supplement musicians in their creative process.}

Recent generation systems have tended to focus on providing the user with some degree of control  over the generation, either through conditioning signals or through priming. For example, \cite{meade2019exploring} and \cite{louie2020novice} both explore a variety of possible ways to condition a generated musical sequence, from composer to histograms of loudness (i.e. MIDI velocities), and \cite{herremans2019morpheus} modifies an existing piece of polyphonic music to fit a musical tension profile, provided by the user.\hide{ and \so{?check anna's?}. MusicTransformer is particularly adept at staying consistent with thematic material presented in a long priming sequence~\cite{huang2018music} (e.g. \so{include link to black key etude?}).

\socheck{maybe: add somewhere * challenges/interest of providing musicians with controls over generated music?\\}}

\subsection{ML-based Tools for Musicians}

\hide{\so{possibly move some of the related work sections to back of the paper e.g. to a longer discussion section}
    * maybe one of dani's posts?\\
}
As more work in machine learning is being done with the goal of providing musicians with tools for supporting music creation, more attention is being paid to how these proposed tools are being used \citep{huang2020ai}.
A motivation for much of the work on controlling musical generation has been that one of the underlying recent goals is to build tools for musicians and producers (of varying skill levels). That is, automating music generation is not an end in itself, but rather it is intended to provide people with new processes to make music themselves. \textit{Piano genie}~\citep{donahue2019} is an extremely entertaining system that allows anybody to play on a small toy keyboard and have it turn into impressive piano music that follows rhythmic and other musical features of the original input. Castro~\citep{castro2019performing} describes and very effectively demonstrates a framework for structured musical improvisation with trained generative models. Others have also explored musical improvisation with a generative model as well, e.g. ~\cite{bretan2017deep,roberts2016interactive}. Following earlier works on timbral conversion~\citep{huang2018timbretron, mor2018universal}, the DDSP project~\citep{engel2020ddsp} provides the user with a sophisticated tool \hide{\so{give more details on how DDSP differs from prev systems}} for transforming timbral and other characteristics of an existing audio recording. For example, the user can record themselves singing and then use DDSP to process that audio keep the pitch and convert it into the sound of a trumpet.

\hide{\socheck{add IUI paper?}}

Our current system is designed in this spirit as well: it is not meant to create a piece of music, but rather to generate musical material, derived from speech and adjusted to fit a musical language model, that a composer can then work with, as we will describe in Section~\ref{ss:use_musicians}.

\subsection{Speech to Music}
\label{s:background-s2m}

There is a long musical tradition of looking to speech itself as a source of inspiration. the great Brazilian musician Hermeto Pascoal called the technique \emph{Som da Aura}~\citep{pascoal2}, and it has been the basis of many songs, videos, and even entire albums \hide{\href{https://www.youtube.com/watch?v=C83_6xwj00A}{songs}, \href{https://www.youtube.com/watch?v=-RQPeoyqyP4}{videos} and even \href{https://www.youtube.com/watch?v=mTpHkMxb89M}{entire albums}}~\citep{spearin}, \hide{\socheck{fix band ref}}in musical contexts of jazz, classical~\citep{reich88different},\hide{or Peter Ablinger's "Deus Cantando"} African~\citep{beier} and other traditions. \cite{danioore} compiled a playlist of over 300 musical videos in response to Donald Trump, many of them using his idiotsyncratic speech patterns as source material for musical compositions. While getting speech samples is easy, ``finding''  coherent melodic structure underlying speech is a challenging task even for experienced musicians. One system providing speech-to-melody functionality is a commercial audio workstation software~\cite{Ableton}, but the automated results of such conversion can be musically complex and perhaps unintuitive both rhythmically and harmonically, since the voice often moves continuously through many notes. That is, when a novice user applies such a conversion, it may be unclear to them what to do with the resulting melody. \hide{An example of this can be heard at the timestamp interval (00:00:56 -- 00:01:09)
\socheck{fix this: need to link to the example where it's the raw conversion, or provide timestamp in the video}
of the sample demo video \footnote{\label{videolink} Please check this link for the demo video: \url{https://youtu.be/IjTnt_MP86M}.}, where a phrase of speech was converted directly to MIDI. } When musicians do this conversion effectively (by ear), they combine the important pitches they hear in the voice \emph{together with extensive musical skill}, perhaps analogously to how speech recognition systems incorporate language models.

One of our contributions here is not just in the automated conversion from speech to pitch, but in the multi-step pipeline to do so: the use of a Transformer allows us to incorporate a musical language model in this way, and the sparsification can work because the musical transformer has been trained to effectively fill in the gaps between such constraints. The Transformer's role here is critical, because it ``makes musical sense'' of the provided constraints, i.e. it chooses notes and rhythms such that those constraints feel musically natural. For example, we have observed the system adding in a note just a fraction of a second before the melody begins, in such a way as to give the resulting filled-in excerpt a much clearer rhythmic structure than it had by the constraints alone\footnote{{\href{https://youtu.be/IjTnt_MP86M?t=151}{Link to demo video}}}. \hide{Indeed, at 00:02:35 in the demo video $^1$, we found it interesting that the transformer added a note just a fraction of a second \emph{before} the given melody begins, helping to give the resulting filled-in excerpt a much clearer rhythmic structure than it did by the constraints alone, which sounded almost a-rhythmic (e.g. at 00:02:31 in demo video $^1$). }

Our approach is unique in its use of Transformers, conditioned on sparse musical note representation of speech, to generate a complete sequence of musical notes following the input speech pattern. The result is a new compositional tool that allows any user---regardless of musical training---to use their own speech to generate musical, satisfying melodies, while still being able to hear the direct connection between their recorded speech and the resulting music.

\subsection{Transformers}
Transformer is a sequence-to-sequence model introduced by \cite{vaswani2017attention} as an improvement over recurrent neural networks. Typically, the sequence-to-sequence model contains an encoder and a decoder, each comprising of a set of self-attention layers. The encoder processes an input sequence and feeds into the decoder, which autoregressively generates the output sequence. In our case, the input and output are MIDI sequences.

Besides the Music Transformer \citep{huang2018music}, our training setup and the Transformer architecture also draws inspiration from the Masked Language Modeling tasks and/or the architecture used in training Transformer models for natural-language understanding and generation tasks. In the Masked Language Modeling (MLM) task, the model is asked to predict a part of the text that is masked out and replaced with mask tokens. BERT~\citep{bert} originally demonstrated the potential of training a Transformer on the MLM task and has directly/indirectly inspired a line of works including T5~\citep{t5}, XLNet~\citep{xlnet}, and BART~\citep{bart}. Architecture-wise, BERT is an encoder-only Transformer and is not suitable for sequence generation out-of-the-box (see \citep{bert_gen}); as described below, we use an encoder-decoder architecture and replace a contiguous sequence of tokens to be masked in our input sequence with a single mask token  -- unlike BERT, which has to replace each word in a span with a mask token on its own, a direct consequence of using an encoder-only architecture -- and train the decoder to predict the masked portions following architectures like T5~\citep{t5}, BART~\citep{bart}, and SpanBERT~\citep{spanbert}.

\section{Contributions}
\label{s:contributions}

\begin{figure}
    \centering
    \includegraphics[width=0.9\textwidth]{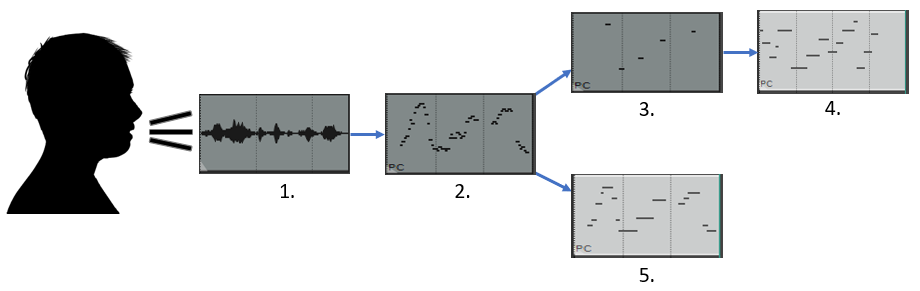}
    \caption{The speech to music conversion pipeline. From the speech audio signal (1), we extract $F_0$, $F_1$, $F_2$, or $F_3$ formant values to obtain a symbolic audio representation (2). From this, we sparsify the sequence according to a heuristic (3) and apply the gap-filling Transformer (4), or alternatively, we directly apply the denoising Transformer (5).}
    \label{fig:pipeline}
\end{figure}

In this work, we demonstrate a system\hide{({\color{blue}{\href{https://www.youtube.com/watch?v=IjTnt_MP86M}{video link}}}) \so{Check video}} that will generate a musical line (melody) that follows the prosody of speech recorded or provided by the user. 
An overview of the system pipeline is shown in Figure~\ref{fig:pipeline}.

Our contributions include the following:
\begin{itemize}
\setlength\itemsep{0pt}
\item a functional pipeline for conversion from speech to music, despite that no such parallel dataset exists (Figure~\ref{fig:pipeline} and Sections~\ref{s:Prepreocess}-\ref{s:evaluation})
\item definition of a new set of musical composition tasks for denoising and gap-filling
\item translation of these tasks into Transformer training
\item exploratory analysis of how the translation adjusts characteristics of the raw musical signal obtained directly from speech (Section~\ref{ss:Q_analysis})
\item an interactive tool that can be used online by anyone (Section~\ref{s:UI})
\item demonstration and discussion of how this tool can be used by expert musicians for creating musical excerpts (Section~\ref{ss:use_musicians}).
\end{itemize}

\section{Pre-Processing Speech Data}
\label{s:Prepreocess}

\subsection{Feature Extraction}

Spectral and prosodic features are commonly used to represent speech signal characteristics.
The features we consider in this work are the following:

\textit{Fundamental frequency $F_0$} refers to the rate of the vocal fold vibration. Several algorithms were proposed to estimate the $F_0$ contour of speech \citep{de2002yin, yegnanarayana2009event, morise2009fast}. In this work, we use the DIO algorithm \citep{morise2009fast, morise2016world} to estimate $F_0$, based on period detection of the vocal fold vibration. DIO is a fast and reliable algorithm for estimating $F_0$ for speech and singing voice.

\textit{Formant frequencies $F_1, F_2$}, and \textit{$F_3$} refer to resonances of the vocal tract system during production of speech \citep{epps97}. Formant estimation from speech is a challenging problem, and algorithms with varying complexity have been proposed for this task~\citep{boersma2001praat, deng2006adaptive, durrieu2013source}. In this work, we use the approach proposed in \cite{boersma2001praat}, which is based on a simple linear prediction analysis \citep{makhoul1975linear, rabiner1993fundamentals}. 

\textit{Loudness contour of the speech signal} is computed as the mean of the perceptually weighted (A-weighting) short-term power spectrum of speech \citep{mermelstein1975automatic}. In this work, we consider the loudness contour to select the regions of interest in speech using two different sparsification techniques.

\subsection{Sparsification techniques}
\label{ss:spars_tech}
The sparsification step is used to select the regions of interest in speech which guide the transformer to generate a musical line (melody) that follows the characteristics of speech provided by the user.
In this paper, we follow two approaches to select the regions of interest in the input speech signal.
1) a heuristic-based approach and
2) a syllable-nuclei-based approach.


\noindent\textit{Heuristic-based approach}: The steps in the heuristic-based approach for sparsification of speech are as follows:
\begin{enumerate}
\item Compute the short-term loudness contour \citep{mermelstein1975automatic} of the speech signal. The short-term loudness contour is extracted using a window of length $50$ ms and a frame-shift/hop-size of $20$ ms.
\item Smoothen the loudness contour using a moving-average smoothing technique.
\item Select only those values in the loudness contour which are higher than a pre-defined threshold and discard other values. The pre-defined threshold is set based on empirical analysis of the loudness contours obtained from a set of speech signals.
\item Consider only the values of $F_0$, $F_1$, $F_2$ and $F_3$ in the regions of interest as obtained in Step-3 for further processing.
\end{enumerate}

\noindent\textit{Syllable-nuclei-based sparsification:} In this approach we consider syllable nuclei \citep{ladefoged2014course} positions in the speech signal as the regions of interest.
Segmenting the loudness contour recursively using the convex-hull algorithm provides a syllable-level segmentation of speech. The peak in the loudness contour in each segment refers to the \textit{syllable nuclei} \citep{zhang2009speech}. In this work we consider the algorithm in \cite{de2009praat} which uses a combination of intensity \citep{pfitzinger1999local} and voicedness \citep{pfau1998estimating} to detect the syllable nuclei locations in speech (Described further in Appendix~\ref{s:appendix-syllable}).



\subsection{Sparsification levels}
We allow three levels of sparsification i.e., low, medium and high, for each of the two sparsification techniques discussed above.
In low-level sparsification, most of the values of $F_0$, $F_1$, $F_2$ and $F_3$ are retained whereas in the high-level sparsification, very few of the $F_0$, $F_1$, $F_2$ and $F_3$ values are retained.
For heuristic-based sparsification, the threshold is varied depending on the level of sparsification. For syllable-nuclei-based sparsification, the context (number of frames selected) around each syllable nuclei is varied depending on the level of sparsification.
For instance in heuristic-based approach, for lower level of sparsification, a lower threshold value is considered. Similarly for the syllable-nuclei-based sparsification, two values on each side of the syllable nuclei (along with the value at syllable nuclei) are retained.

\begin{figure}
    \centering
    \includegraphics[width=0.64\textwidth]{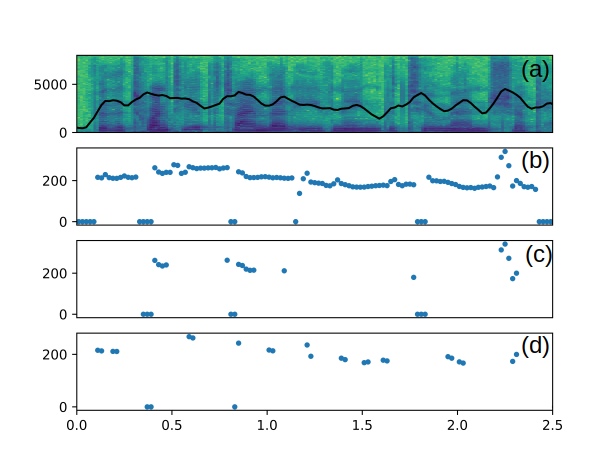}
    \caption{Figure depicting the sparsification techniques. (a) spectrogram of input speech signal. The black line shows the  smoothed loudness contour, (b) $F_0$ contour of the speech signal, (c) $F_0$ values retained after applying heuristic-based sparsification and (d) $F_0$ values retained after applying syllable-nuclei-based sparsification}
    \label{fig:sparsification}
\end{figure}

Figure~\ref{fig:sparsification} shows the retained $F_0$ in the regions of interest as obtained by the two sparsification approaches. For Figure~\ref{fig:sparsification}, we considered medium-level of sparsification for both approaches.

\section{MAESTRO Dataset}

\begin{figure}
    \centering
    \includegraphics[width=0.65\textwidth]{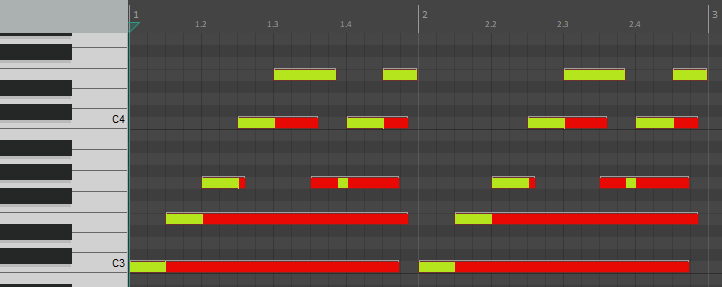}
    \caption{An example of the skyline heuristic on a polyphonic file. Green indicates the monophonic sequence that is extracted from this file.}
    \label{fig:skyline}
\end{figure}

As mentioned, we used the MIDI format to obtain tokenized representations of music, which captures note pitch (integer in [0-127]), velocity, start time, and end time. All models were trained using the MIDI component of the MAESTRO V2.0.0 dataset ~\citep{hawthorne2018enabling}. This dataset consists of approximately 200 hours of professional classical piano performances. Because the extracted speech sequences from voice recordings are monophonic (single-pitched at any given time), our generative models were entirely focused on monophonic generation. In order to use MAESTRO, which consists of polyphonic files, we first pre-processed by extracting the highest note being played at any given time, in a so-called ``skyline'' heuristic (see Figure~\ref{fig:skyline} for an example). Note that this method is far from being proper melody extraction, which in itself, is a challenging problem in the music domain. Rather, this method provides us with harmonically reasonable sequences of monophonic music.

To provide numerical inputs to the models, we represented monophonic sequences as a time-series of held pitches with a time-step of $20$ ms, which is around the level of human imperceptibility. For example, an input of:
\[
    [S,40,40,40,40,40,42,42,42,42,42,0,0,0,0,0,E],
\]
indicates the 40th piano note being held for $0.1$ s, followed by the 42nd note being held for $0.1$ s. $S$ and $E$ are the start and end tokens respectively, and $0$ represents silence. Due to the space complexity limitations of Transformer, we restricted inputs to a length of $502$ ($10$ s sequence with start and end tokens), and we prepared $10$ s samples from MAESTRO accordingly.

\section{Seq2seq Methods}

\subsection{Overview}

We trained sequence-to-sequence models that would modify raw speech sequences by injecting musical structure learned from MAESTRO. We settled on two approaches:

\begin{enumerate}
    \item Apply one of the two sparsification techniques to decrease the density of the raw speech sequence and fill the gaps back in with a ``gap-filling'' Transformer model trained on MAESTRO.

    \item Directly apply a denoising Transformer model to the raw speech sequences that is trained to reduce the chromaticism of the sequence in a way that reveals the underlying musical structure.
\end{enumerate}

\subsection{Gap-filling Transformer}

The gap-filling model consists of an encoder-decoder architecture ~\citep{vaswani2017attention}. The inputs to the encoder and decoder are both initially passed through a $512$-dimensional embedding layer which is learned during training. Each encoder and decoder layer has a self-attention component and a feed-forward component. The encoder and decoder both have $6$ layers, each with $8$ attention heads and $1024$-dimensional feed-forward layers. Observing that the note sequences consist of repeated tokens, we experimented with predicting the token-pitch and its token-count in the same unrolling step; empirically, we found this scheme to work better than predicting the same token-pitch over successive unrolling steps. Therefore, on a given encoder-decoder input pair, the output of the decoder is passed to both a linear pitch predictor, and a linear-sigmoid token count predictor. During training, counts are scaled from $[1,500]$ to $[0,1]$, to suit the sigmoid function range.

In addition, we also employed the use of relative position-based attention, following the method of~\cite{huang2018music}, which has been shown to perform better than sinusoidal positional encoding at capturing long-term structure. 

We designed a task where we mask segments of a MAESTRO sequence  and learn to map masked sequences to the original sequence. Random masks are created by the following procedure:
\begin{enumerate}
\item Construct a multiset $M$ from the set of token lengths $S = \{25, 50, \dots, 150\}$ such that $\sum_{i \in M} i = 150$.
\item For every element $i \in M$, $i$ consecutive tokens are replaced with the $<$gap$>$ token in the input. This is done by uniformly choosing a continuous segment of length $i$ that does not contain any $<$gap$>$ token.
\end{enumerate}
The decoder is trained to predict what the gap tokens should be and therefore, the loss is evaluated only over the predictions that correspond to the $<$gap$>$ tokens: the pitch prediction is evaluated with a negative log-likelihood loss, while the token count prediction is evaluated with mean squared error loss. Note that this is reminiscent of the CocoNet project~\citep{huang2017counterpoint} in which polyphonic musical scores are ``filled in''. Additionally, we augment the dataset by performing pitch transposition between $-5$ and $+5$ semitones during sampling. Adam optimizer was used, following the learning rate schedule described in~\cite{vaswani2017attention}. A dropout value of $0.1$ was used on all layers during training. The model was trained with a batch size of $8$ for approximately $230,000$ iterations on $4$ Nvidia Tesla P100 GPUs.

\subsection{Denoising Transformer}

The denoising model also follows an encoder-decoder structure, almost identical to the gap-filling model. The embedding dimension is $128$, followed by $2$ encoder and decoder layers, with $1024$-dimensional feed-forward sub-layers. Each attention module has $2$ attention heads. The decoder output is passed to a linear layer to do pitch prediction only. Unlike the gap-filling model, pitch is predicted token by token.

The training task was to map noisy inputs to their original state. Noisy inputs are created by the following procedure:
\begin{enumerate}
    \item Generate a sequence of random variables from a normal distribution of zero mean and unit variance, with a length equal to the input.
    \item Round the random variables to the closest integer.
    \item Add the sequence of random variables to the input sequence under the condition that $0$ tokens remain, tokens in the range $[1,88]$ stay in this range, and end tokens remain the same.
\end{enumerate}
Pitch prediction is evaluated under a negative log-likelihood loss. The optimizer and learning rate schedule are identical to \cite{vaswani2017attention}. A dropout value of $0.1$ was used on all layers during training. The model was trained with a batch size of $32$ for approximately $260,000$ iterations on a single Nvidia Tesla P100.

\subsection{Inference}

As described above, the MIDI representation of a melody consists of a sequence of note-pitches along with their velocities, and start and end times. The Transformer models determine the note-pitches and their durations in the output sequence, while their velocities are determined from the speech loudness by mapping the loudness to an integer in the closed-interval [0,127]. The inference is done depending on the Transformer model as described below:
\begin{itemize}
    \item Gap-filling Transformer: Either one of the $F_0$, $F_1$, $F_2$ or $F_3$ sequences, extracted from the input speech, is sparsified using any one of the techniques described in Section~\ref{ss:spars_tech}. The parts of the sequence chosen to be discarded as a part of the sparsification process are replaced with $<$gap$>$ tokens. The pitch is randomly chosen from the softmax output of the model, while the length is deterministically obtained by mapping the sigmoid output back to [1,500]. The Transformer model is then continuously unrolled unless all of the gap tokens are filled; if the token-count predicted in the latest unrolling step does not match with the remaining number of consecutive gap tokens, the extra tokens are discarded.
    \item Denoising Transformer: Either one of the $F_0$, $F_1$, $F_2$ or $F_3$ sequences, extracted from the input speech, is provided as input to the denoising Transformer, and the resulting denoised output -- of the same length as that of the input -- is used to infer the pitches and durations of the refined MIDI sample.
\end{itemize}



\section{User Interface}
\label{s:UI}

\begin{figure}[ht]
    \centering
    \frame{\resizebox{\linewidth}{!}{
    \includegraphics{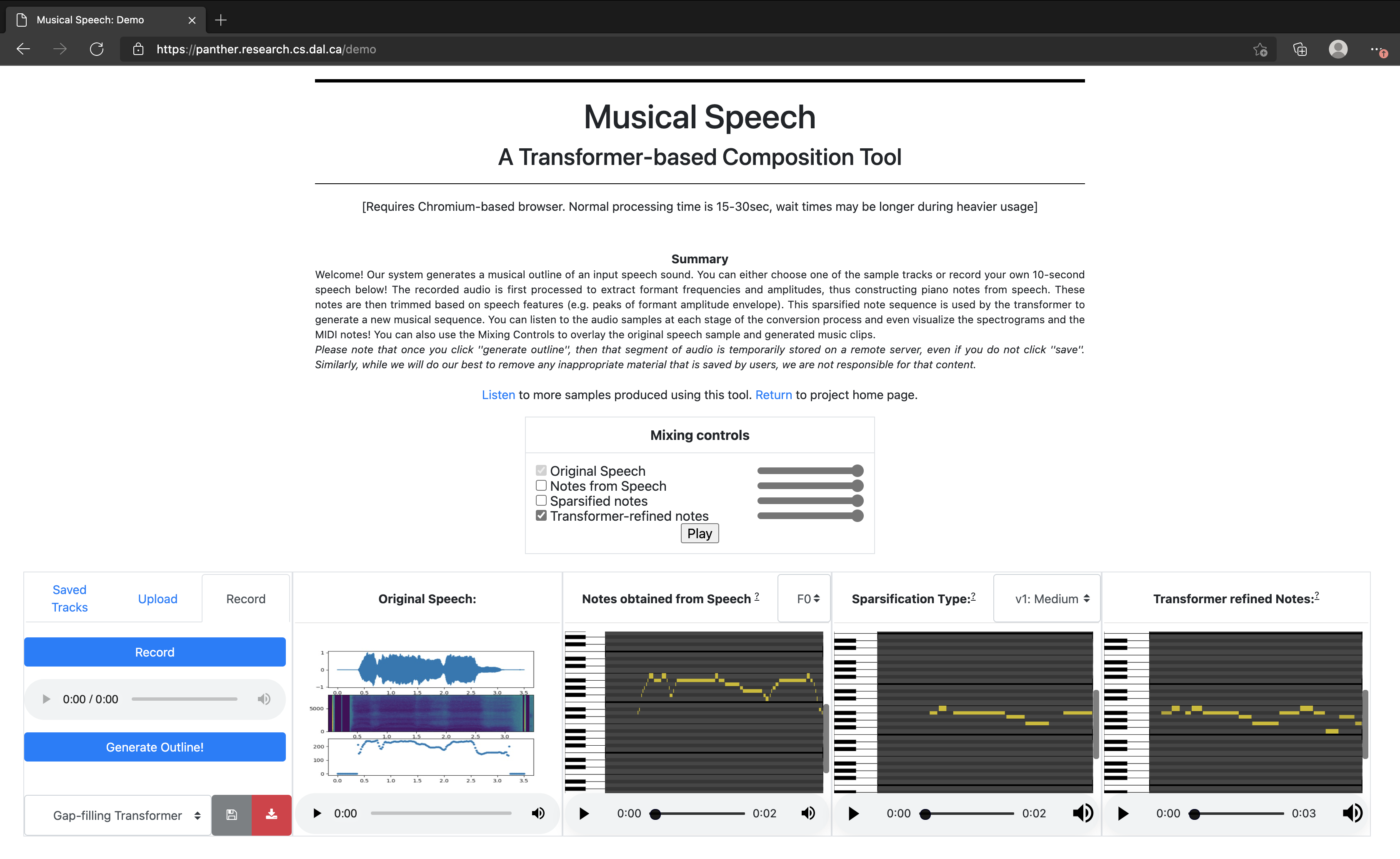}}}
    \caption{The web interface for interacting with the musical speech system. }
    \label{fig:screen_capture}
\end{figure}

Figure~\ref{fig:screen_capture} shows the user interface of the online interactive version of the system, which enables the following functionality:
\begin{enumerate}
\setlength\itemsep{0.1em}
\item The UI allows for the composers to either record their live speech using a microphone or to upload a pre-recorded audio file.
\item The chosen audio file will be processed as described above, with all the intermediate results made available in the MIDI format as shown.
\item Depending on the speech sample and the kind of inspiration that the composers are looking for, certain configurations may be more preferred; also keeping in mind that it is not easy to create a one-size-fits-all pipeline, we allow the users to choose the type of model, the formant frequencies to extract, and the sparsification method to choose.
\item The composers can interactively mix input speech, the intermediate MIDI results, and the generated final MIDI sample while the mixed output plays on a loop until they identify the desired proportions.
\item In order to use the generated samples in their compositions, the composers could choose to either download the MIDI samples corresponding to a particular configuration or all possible configurations.
\end{enumerate}

\section{Evaluation}
\label{s:evaluation}

Creativity support tools have been referred to as a ``Grand Challenge for HCI Researchers'' \citep{shneiderman2009creativity}. A significant part of this challenge, especially as the tools themselves are becoming more powerful, is also in their evaluation. Recently, \citet{remy2020evaluating} surveyed over 100 papers presenting creativity support tools, with the focus of how to evaluate such tools. They distinguish, for example, between evaluating usability and creativity: ``traditional usability methods emphasize aspects such as efficiency, precision, error prevention, and adherence to standards [..], but do not address core dynamics of creative work, such as exploration, experimentation, and deliberate transgression of standards''. Arguably, the latter considerations are indeed far more relevant in the present work than the former ones. While Remy et al indicate that the task of evaluating creativity support tools is very much an open problem, they do propose a set of recommendations on how to evaluate them in future. These recommendations include recruiting domain experts (in this case, professional musicians) and  considering longitudinal, in-situ studies. (41\% of the evaluations they surveyed lasted less than or up to one hour, and altogether 82\% lasted less than or up to a day; neither of these would generally qualify as longitudinal).

Recognizing, then, that a rigorous such evaluation is outside the scope of this paper, we present two related analyses. First, in Section~\ref{ss:Q_analysis}, a quantitative analysis briefly examines the question of whether the musical language modeling aspect of our tool is doing what it purports to be doing, e.g. does it shift the statistics of speech-based sequences towards those of musical sequences?

Second, in Section~\ref{ss:use_musicians}, we present a few samples created by professional musicians, one of whom spent many hours across several weeks experimenting with the system.  Many more samples were created during that time; we choose just a small and diverse set here. We also note that some of the design choices made were in fact done so collaboratively with the expert musician, by providing them with early versions of the tool and incorporating their feedback (a well-known approach for effective design processes~\citep{rogers2011interaction}).

\subsection{Quantitative Analysis}
\label{ss:Q_analysis}

\begin{figure}
    \centering
    \includegraphics[scale=0.830625]{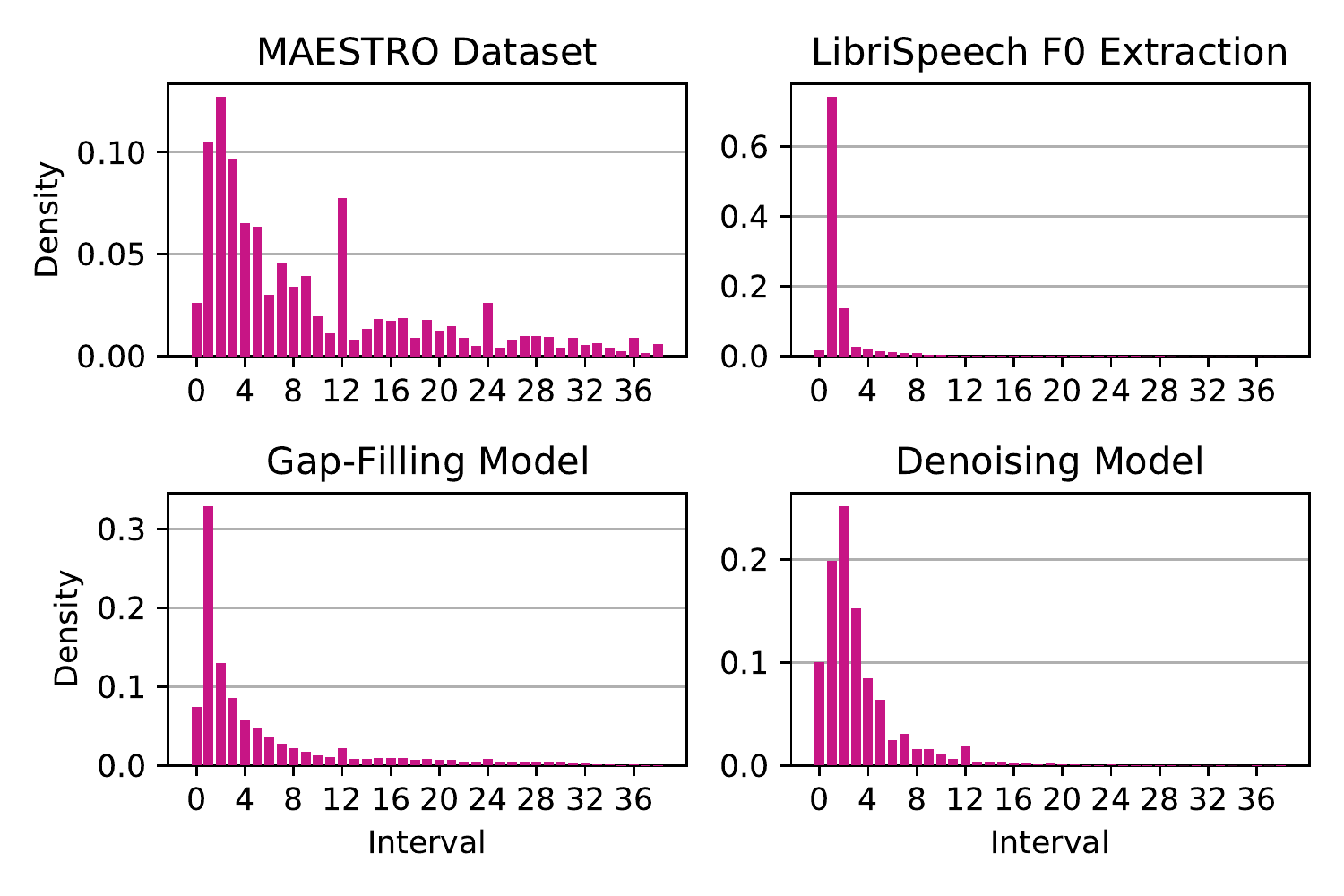}
    \caption{Frequency of musical intervals in the monophonic MAESTRO dataset, the $F_0$ values extracted from LibriSpeech samples, and the outputs of the gap-filling and denoising model respectively.}
    \label{fig:intervals}
\end{figure}

To evaluate the effectiveness of our models, we compared statistical features of the MAESTRO dataset against the Transformer outputs. In these experiments, we used the test-clean partition of the LibriSpeech dataset~\citep{panayotov2015librispeech} as a source of voice recordings. In particular, we looked at the distribution of the intervals between adjacent pitches in sequences. Figure~\ref{fig:intervals} shows the intervallic distributions for the skyline of the MAESTRO dataset, the extracted $F_0$ sequences, the outputs of the gap-filling model with medium-level syllable-nuclei sparsification applied, and the outputs of the denoising model.

The distribution of intervals from the MAESTRO dataset are indicative of common patterns in classical music. For example, small intervals of 1, 2, or 3 semitones are the most common kind of musical movement. An interval of 6 semitones is uncommon relative to 5 or 7 semitones (in music terms, a tritone, a perfect 4th, and a perfect 5th respectively), whereas 12 semitones is common relative to 11 or 13 semitones (an octave, a major 7th, and a minor 9th respectively). The shape of the distribution is roughly periodic with a decreasing scale, with a period of 12 semitones.

Extracted $F_0$ values from speech are highly chromatic (intervals of 1 semitone are frequent) and this is reflected in the distribution for LibriSpeech $F_0$ speech sequences. However, the distributions for the model outputs lie somewhere in the middle of the spectrum between pure music data and extracted $F_0$ values. The pipeline utilizing the gap-filling model maintains portions of the chromatic $F_0$ sequence, but noticeably increases the occurrence of larger interval sizes. The denoising model is not constrained to keep any part of the $F_0$ sequences, resulting in a distribution closer to that of the MAESTRO sequences.

\subsection{How is this used by musicians?}
\label{ss:use_musicians}

To examine how this system is used by musicians, we provide a set examples of different musical excerpts created with the system illustrating a set of use cases. Note that for some of these examples, listening with headphones is strongly recommended.

\subsubsection{Use Case 1 - A Fun Tool for Direct Speech to Melody}

\textbf{Fun to have fun} (\textcolor{blue}{\href{https://jasondeon.github.io/musicalSpeech/workshop_samples/funToHaveFunExpo.wav}{[link]}}): This example demonstrates a musician using the system to sketch out an accompaniment for a speech recording of a passage from a children's book~\cite{geisel1957cat}. First we hear the recording on its own, followed by the dense sequence of $F_0$ values (0:07). The denoising Transformer rewrote the extracted frequencies into a new melody (0:14). The musician added a simple rhythm section (0:21), and adjusted the melody slightly, still roughly aligned harmonically and rhythmically with the original voice (0:27). The result is a melody, harmonic, and rhythmic accompaniment that very closely follows and ``supports'' the contour of the original spoken voice recording.

\subsubsection{Use Case 2 - Direct Speech to Sound}

\textbf{Hi. I'm a robot. I love you.}  (\textcolor{blue}{\href{https://jasondeon.github.io/musicalSpeech/workshop_samples/robot_part1.m4a}{[Part 1 Link]}}): In this example, a musician uses all of the formants to create sounds that closely match the sounds of the original words. The format of this audio file is 3 subsequences, each subsequence itself consisting of 3 parts: (speech) (formantV1) (formantV2). That is, (formantV1) uses a musical synthesizer to render the full set of formants corresponding to (speech). (formantV2) is exactly the same, but rendered using a different musical synthesizer. The clip consists of 3 such subsequences. This gives us a sense of the range of sounds that can be produced using a single piece of text, and the impact of choice of instrument used.
\vspace{5mm} 

\noindent \textbf{Hi. I'm a robot. I love you.}  (\textcolor{blue}{\href{https://jasondeon.github.io/musicalSpeech/workshop_samples/robot_part2.mp3}{[Part 2 Link]}}): In this one, the same speech was used, and all of the musical material was generated by the gap-filling (V2) music-to-speech system using the F0 output, and then orchestrated and mixed by a musician so that it ``worked'' as a short musical clip. It was noted that while the system generated the material quickly, it was still a slow, careful process for the musician to work with that raw material to bring it to this point.

\subsubsection{Use Case 3 - Musical Material Generator}
\label{s:usecase3}

\textbf{Moo-cow}
(\textcolor{blue}{\href{https://jasondeon.github.io/musicalSpeech/workshop_samples/moo cow.wav}{link}}): A musician created this piece by incorporating the MIDI files our system generated from a ten-second spoken voice recording (itself also included throughout the piece). The singing is the only audio recording added other than the original voice and the generated MIDI files. That is, \emph{all raw MIDI files} (i.e. all non-vocal clips used for instruments and synths) in this piece were generated by our system. They were then assembled, orchestrated, and mixed to create and produce this resulting 30-second excerpt. To provide insight into the musical process, Appendix~\ref{s:appendix-score} shows some parts of the score of the track as it appeared on the musician's digital workstation software.
\vspace{5mm} 

\textbf{Thing 1 and Thing 2}
(\textcolor{blue}{\href{https://jasondeon.github.io/musicalSpeech/workshop_samples/thing one and thing two.wav}{link}}): This is another example similar to the previous, where a musician has arranged and orchestrated the original speech recording and MIDI files to create the track, with all raw MIDI files generated by the system.

\section{Conclusion}
We have proposed a pipeline for translating speech into musical building blocks as an interactive tool for musical composition and illustrated its effectiveness through examples of music created with the help of our tool. In achieving this, we make novel use of speech-processing techniques
to effectively sidestep the need for a paired speech-music dataset.

Our evaluation reveals that different training objectives and architectural choices can give rise to different forms of dependence on the conditioning input. We use a set of created musical excerpts to demonstrate aspects of the system and examples of how it can be effectively used to generate new musical pieces.


\section{Acknowledgements}
We thank Rudolf Uher and group for insightful discussions. We thank the Canadian Institute for Advanced Research (CIFAR) for their support. Resources used in preparing this research were provided, in part, by NSERC, the Province of Ontario, the Government of Canada through CIFAR, and companies sponsoring the Vector Institute \url{www.vectorinstitute.ai/#partners}. We thank the Magenta team at Google for making the MAESTRO dataset publicly available.


\bibliography{main.bib}

\newpage
\appendix

\section{Sample Elements of a Composition}
\label{s:appendix-score}

\begin{figure}
    \centering
    \includegraphics[width=0.8\textwidth]{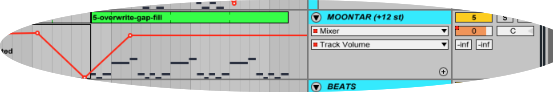}
    \caption{MIDI files labeled ``overwrite-gap-fill'' (green), trigger a moon guitar [aka yueqin] (labeled ``MOONTAR'')}
    \label{fig:moontar}
\end{figure}

\begin{figure}
    \centering
    \includegraphics[width=0.8\textwidth]{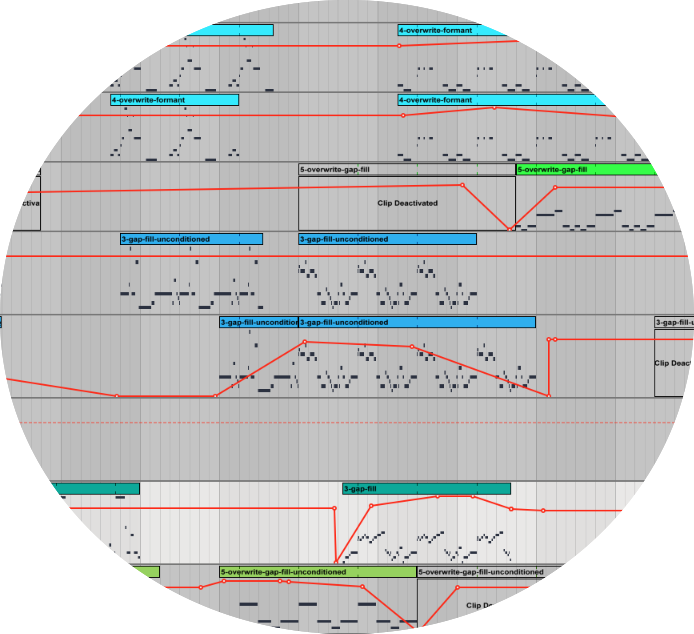}
    \caption{The MIDI files (green– and blue-tabbed files containing tiny black rectangles of different heights and lengths representing pitch and duration information respectively, and which trigger the MIDI instruments in the tracks that they populate) are each labeled along their green and blue tabs according to the manner in which the given raw MIDI file was generated.\hide{, e.g.: \so{colour the text as dani did}}
}
    \label{fig:midi-files}
\end{figure}

Figures~\ref{fig:moontar}~and~\ref{fig:midi-files} illustrate some elements of the \textit{Moocow} composition described in Section~\ref{s:usecase3}.

\hide{
\begin{figure}
    \centering
    \includegraphics[width=0.9\textwidth]{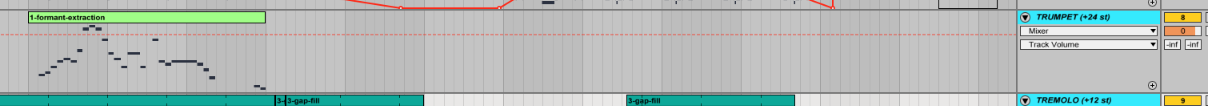}
    \caption{A MIDI file labeled ``formant-extraction'' (green), triggers a TRUMPET (track 8), e.g.:}
    \label{fig:trumpet}
\end{figure}
}

\hide{
\begin{figure}
    \centering
    \includegraphics[width=0.9\textwidth]{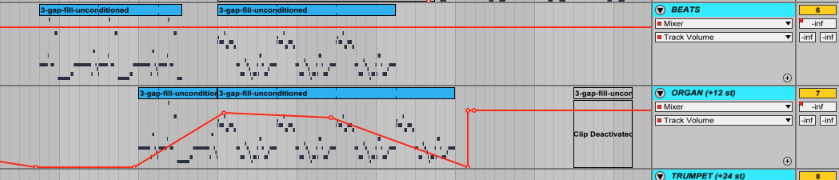}
    \caption{MIDI files labeled ``gap-fill-unconditioned'' (blue), trigger an ORGAN (track 7) and electronic drum BEATS (track 6)}
    \label{fig:beats-and-organ}
\end{figure}
}

\section{Syllable-nuclei detection}
\label{s:appendix-syllable}

A brief description of the steps to locate syllable nuclei positions in speech is as follows:
\begin{enumerate}
\setlength\itemsep{0.5em}
\item Extract the loudness contour from the speech signal. The extracted loudness contour is Smoothed using moving-average smoothening technique.
\item All peaks in the smoothed loudness contour are considered as the initial set of potential syllable nuclei.
\item Discard all peaks below a pre-fixed threshold. Here we set this threshold to be 2 dB above the median loudness measured over the entire input speech recording.
\item Inspect the preceding dip in intensity of loudness contour for each peak. Consider only peaks with a preceding dip of at least 2dB with respect to the current peak as a potential syllable nuclei and discard other peaks.
\item Extract the $F_0$ contour to obtain the voiced/unvoiced regions in the speech signal. Regions with $F_0$ values below $40$ are considered as unvoiced regions. Discard peaks within the unvoiced regions.
\item Peaks retained after step-4 are considered as syllable nuclei.
\end{enumerate}

\end{document}